\documentclass[aps,prl,twocolumn,groupedaddress,superscriptaddress,showpacs]{revtex4-1} 
\usepackage{graphicx} 
\usepackage{color} 
\usepackage{amsmath} 
\usepackage{amsfonts} 
\usepackage{amssymb} 
\usepackage{dcolumn} 
\begin{document} 
\title{Loop erased random walk on percolation cluster: Crossover from Euclidean to fractal geometry} 
\author{E. Daryaei} 
\email{edaryayi@gmail.com} 
\affiliation{Department of Physics, Sharif University of Technology, Tehran, P.O.Box: 11365-9161, Iran} 
\affiliation{Faculty of Science, Neyshabur University, Neyshabur, P. O. Box 91136-899, Iran}  
\author{S. Rouhani} 
\affiliation{Department of Physics, Sharif University of Technology, Tehran, P.O.Box: 11365-9161, Iran} 

\begin{abstract} 
We study loop erased random walk (LERW) on the percolation cluster, with occupation probability $p\geq p_c$, in two and three dimensions. We find that the fractal dimensions of LERW$_p$ is close to normal LERW in Euclidean lattice, for all $p>p_c$. However our results reveal that LERW on critical incipient percolation clusters is fractal with $d_{f}=1.217\pm0.0015$ for $d=2$ and $1.44\pm0.03$ for $d=3$, independent of the coordination number of the lattice. These values are consistent with the known values for optimal path exponents in strongly disordered media. We investigate how the behavior of the LERW$_p$ crosses over from {\it Euclidean} to {\it fractal} geometry by gradually decreasing the value of the parameter $p$ from 1 to $p_c$. For finite systems, two crossover exponents and a scaling relation can be derived. This work opens up a new theoretical window regarding diffusion process on fractal and random landscapes. 
\end{abstract} 

\maketitle 
Diffusion process in disordered media is {\it anomalous}, i.e. mean square displacement (MSD) of the diffusing species has a non-linear relationship with time, in contrast to diffusion on Euclidean lattices, where their MSD is proportional to time in all dimensions \cite{Sahimi12,*Bouchaud90,*Havlin87}. Such disordered media is typically simulated through percolation systems, diffusion on percolation clusters have been studied in great detail \cite{Gefen83,Avraham00}. One could restrict the diffusion of a simple random walk (RW) to the incipient infinite cluster; in this case finite-sized clusters are irrelevant. It is known that, above criticality, $ p>p_c $, diffusion is anomalous over short distances and normal over long distances \cite{Avraham00}. As the occupation probability approaches the percolation threshold, diffusion becomes more anomalous over longer distances. Diffusion on critical incipient percolation clusters is anomalous on all length scales. 
On the other hand, one could erase the loops from trajectory of the RW chronologically, this operation results in the loop erased random walk (LERW) \cite{Lawler80}. This model is equivalent to the classical uniform spanning trees(UST) \cite{Wilson96}, the q-state Potts model in the limit $q\to0$ \cite{LERW_exact}, and the avalanche frontier in Abelian sandpile model (ASM) \cite{Majumdar92}. It is known that the fractal dimension of LERW in $D=2$ is $5/4$ and the upper critical dimension for LERW is $D=4$, with $d_f= 2$ for $D\geqslant4$. Although scaling and universality class of LERW in integer lattice is known, the universality class of this model in fractal landscape and especially in critical percolation had not been hitherto studied. 
\begin{figure}[!htb] 
\includegraphics[scale=0.407]{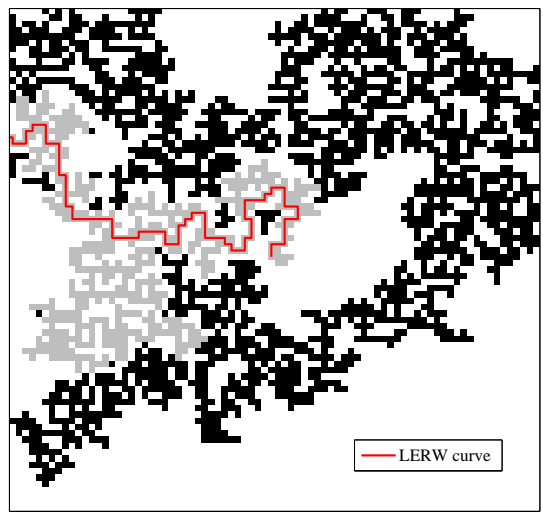} 
\caption{\label{fig1} 
(Color online) The generation process of a 2D LERW on critical percolation cluster on a $80 \times80$ lattice. A random walk starts at middle of the lattice in critical large cluster (shown in black color) and diffuses to outer boundaries; the visited sites are shown in gray, then LERW curve (shown in red) can be obtained by erasing the loops chronologically. 
} 
\end{figure} 
In this paper we study the LERW on percolation cluster, with occupation probability above and equal to the critical value, $p\geq p_c$. 
Our results show that for all $p>p_c$, the scaling behavior of obtained LERW$_p$ curves is close to exact results for LERW on Euclidean lattices \cite{Majumdar92}. However, our results reveal that scaling behavior of this model near critical percolation is completely different, with fractal dimension $d_{f} \approx 1.22$ in 2D and $d_{f} \approx 1.44$ in 3D. Surprisingly, these values are close to a family of curves appearing in different contexts such as, e.g., polymers in strongly disordered media \cite{Porto97}, watershed of random landscapes \cite{Fehr09,Fehr11,Fehr11_2}, bridge percolation \cite{Schrenk12}, and optimal path cracks \cite{Andrade09}. A crossover from {\it Euclidean} to {\it fractal} geometry can be observed by decreasing the value of the parameter $p$ from 1 to $p_c$. To investigate how the behavior of the LERW$_p$ crosses over from between these two universality classes, we have considered the mean total length of LERW$_p$ as a homogeneous function on the lattice size and occupation probability. For the finite systems, two crossover exponents and a scaling relation can be derived. Our results for crossover regime demonstrate that for a fixed lattice size, $L$, there are three distinct scaling regimes, as it has been reported in similar systems \cite{Moreira12,Andrade09}. These regimes are separated by two crossovers. Finally there is a scaling relation between the corresponding crossover exponents and the fractal dimension of LERW.

\emph{Simulation details}.~We start by constructing a porous landscape, we simulate site percolation model on a lattice of size $L$ with free boundary conditions. We are especially interested in studying diffusion properties in single percolation clusters with the fraction of occupied sites $p\geqslant p_c$, the single cluster containing the middle point of the lattice, i.e., $L/2$, is generated in the same way as algorithm particularly \cite{Leath_algorithm}. If the obtained cluster is large enough to connect to at least one of the outer edges, we accept it otherwise we simply ignore it and produce another one. Once a large cluster is obtained, we start diffusion process by setting a RW on middle of the lattice, i.e., $L/2$, and stop when it touches the outer edges for the first time. The LERW curve can be obtained by erasing the loops choronologically. In Fig.~\ref{fig1}, a sample of LERW on critical percolation, $p=p_c$, is shown. 
In planar LERW, simulations on critical percolation clusters were performed on square lattice of size \mbox{$L=2^{n+4}, n=1,2,..,8$}. The number of samples generated for each lattice size ranges from $4\times10^6$ for the smallest system sizes till about $10^4$ for the largest system sizes. 
Moreover, to study this model in higher spatial dimensions, we simulate the same problem on a simple-cubic lattice. 
\begin{figure}[h] 
\includegraphics[scale=0.42]{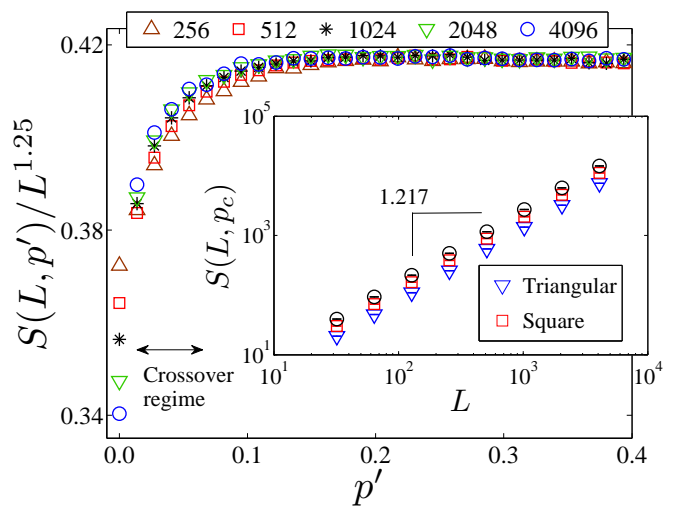} 
\caption{\label{fig::fig2} 
(Color online) Dependence of the mean total length normalized with system size, $S(L,p)/L^{1.25}$, on $p^{\prime}$ for different lattice sizes. Above criticality, LERW$_p$ behaves as expected for Euclidean lattices. However, near the criticality, $S(L,p_c)$ grows slower than $L^{\frac{5}{4}}$. A crossover between two regimes can be observed for $p$ in the neighborhood of $p_c$. \textbf{Inset}: Mean total length of LERW on critical percolation as a function of lattice size $L$ for three different lattices. (the statistical error bars are shown, but are quite shorter and appear as horizontal lines). The slope in the log-log plot corresponds to $d_f=1.217\pm 0.0015$.} 
\end{figure}

\begin{figure} 
\begin{center} 
\includegraphics[scale=0.62]{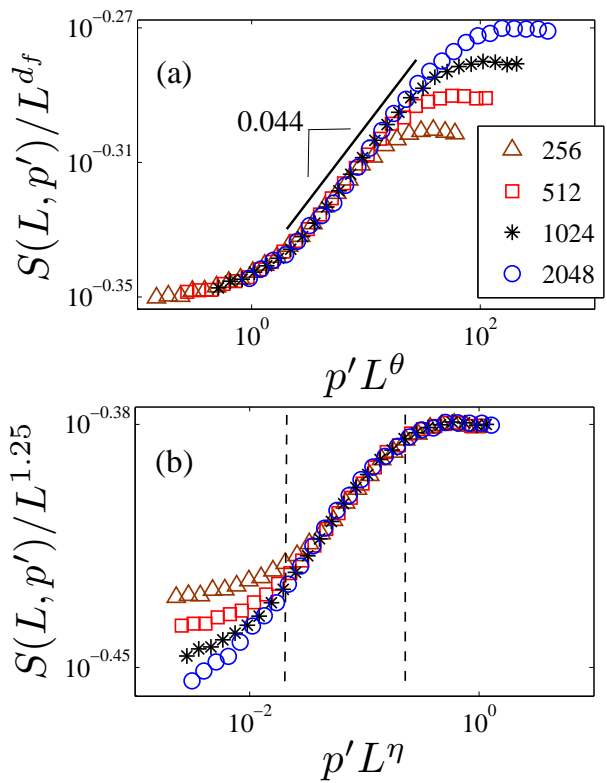} 
\end{center} 
\caption{ 
Crossover scaling and data collapse for LERW$_p$ in different system sizes. 
\textbf{(a)}: Rescaled mean total length $S/L^{d_f}$, where $d_f=1.217$, versus $p^{\prime}L^\theta$ for different system sizes. The scaling function given by equation~(\ref{eq::scaling1}) is applied, with $\theta=0.90\pm0.05$. 
\textbf{(b)}: Rescaled mean total length $S/L^{1.25}$ versus $p^{\prime}L^\eta$, with $\eta=0.15\pm0.02$, for different system sizes. For each finite lattice size $L$, there are three regimes separated by two crossovers. These crossovers scale with two crossover exponents i.e. $\theta$ and $\eta$. By collapsing the data of different lattice sizes in the intermediate regime, a more precise estimate for $\beta$ can be obtained, which is $\beta=0.044\pm0.002$. 
All results have been averaged over $4\times10^4$ samples. 
\label{fig::fig3}} 
\end{figure}

\emph {The fractal dimension}.~We estimate the fractal dimension for the obtained LERW$_p$ by computing the mean total length $S$ for different lattice size $L$, comparing it with $S\sim L^{d_f} $. In the case of normal LERW ($p=1$), the total length of the curves increases with system size as $S(L)\sim L^{\frac{5}{4}}$, consistent with the fractal dimension of Euclidean LERW. By decreasing the occupation probability, $p$, the fractal dimension of these random curves remains unchanged. At percolation threshold, these curves are smoother than normal LERW, and the mean total length diverges with system size with different exponent $S(L)\sim L^{d_f}$, with $d_f\approx 1.22$. Fig. \ref{fig::fig2} shows dependence of the $S(L)/L^{\frac{5}{4}}$ on $p^{\prime}$ for different system sizes, where $p^{\prime}$ is $p-p_c$. The overlap of the different curves confirms that the fractal dimension of the LERW$_p$ above $p_c$ is $5/4$. A small deviation is observed due to finite-size effects. There is a crossover between two different regimes near critical point $p\gtrsim p_c$ which can be observed in Fig.~\ref{fig::fig3}. 
The mean total length of LERW on critical percolation for different lattice size $L$ is shown in the inset of Fig.~\ref{fig::fig3}. We obtain $d_{f}=1.217\pm0.0015$. In order to check universality of this exponent and to show that it does not vary with the coordination number of the lattices, we perform same simulations on the triangular and the honeycomb lattices. We found the same result (see inset of Fig.~\ref{fig::fig3}), which provides strong evidence for the universality of this exponent. 

\emph {Crossover scaling function}.~ As shown in the Fig.~\ref{fig::fig2}, the mean total length of LERW$_p$ increases with increasing occupation probability. For large systems, the mean total length of LERW$_p$ grows with $p$, such that, $S(L,p)\sim p^{\prime\beta}$, where $\beta\approx0.04$ is a novel exponent, which we call length-growth exponent. 
It is known that the anomalous diffusion in percolation clusters occurs only within the correlation length \cite{Avraham00}. At high occupation probability, the correlation length is so small, therefore the mean length of LERW$_p$ can be described as a LERW in Euclidean geometry. When the occupation probability is reduced, the correlation length increases as $p^{\prime 1/\nu}$ diverging at $p_c$, where $\nu$ is the correlation length exponent of the percolation model, with $\nu=4/3$ in 2D, so the system becomes a self-similar random fractal leading to a new universality class. There is a crossover behavior, as depicted in Fig.~\ref{fig::fig2}, from Euclidean to fractal geometry. For the complete crossover scaling of the mean length, $S$ can be considered as a homogeneous function on the relevant scaling fields, $S\left(bL,b^{y_p}p^{\prime}\right)=b^{-y_s}S\left(L,p^{\prime}\right)$ where $b$ is a scaling parameter and $y_s$ and $y_p$ are relevant exponents for $S$ and $p$ scaling parameters respectively. One could restrict attention to the $p\rightarrow p_c$ regime, then for finite size of $L$, it is expected that $S$ scales with $d_f$, so in this regime $y_s=d_f$. The next exponent can be found by trying to collapse the data (setting $b=L^{-1}$). The scaling {\it ansatz} for the mean total length is given by, 
\begin{equation} 
S\left(L,p^{\prime}\right)=L^{d_f} \mathcal{G}\left[p^{\prime}L^{\theta}\right] \ \ , \label{eq::scaling1} 
\end{equation} 
\noindent where $\mathcal{G}\left[u\right]$ is a scaling function, such that, $\mathcal{G}\left[u\right]\sim u^{\beta}$ for small values of $u$, and is nonzero at $u\rightarrow0$. The exponent $\theta=y_p$ is the crossover exponent in the $p\rightarrow p_c$ regime. 
Fig.~\ref{fig::fig3}(a) shows crossover scaling for different lattice sizes, close to the critical point. 
As it shown, we have a good data collapse for small values of $u$ with $\theta=0.90\pm0.05$. For each finite lattice size $L$, there is a crossover point such as $p_{\times_1}^{\prime}$ scales like $L^{-\theta}$, which for $u \ll1$ we have a saturation regime, and for $u\gg1$ results are consistent with $u^{\beta}$ for all lattice size $L$. However for large values of $p^{\prime}L^{\theta}$, we don't observe data collapse and the mean total length behaves as $L^{1.25-d_f}$. 
On the other hand, one could look large values of $p$, it is expected that the mean total length scales with fractal dimension of Euclidean geometry, so $d_s=\frac{5}{4}$ in this regime. If we follow the same strategy as above, we could find another scaling function; 
\begin{equation}\label{eq::scaling} 
S(L,p^{\prime})=L^{\frac{5}{4}}\mathcal{F}\left[p^{\prime}L^\eta\right], 
\end{equation} 
where the scaling function $\mathcal{F}[x]$ have a saturation regime for large values of $x$, and the exponent $\eta=y_p$ is the corresponding crossover exponent in this regime. In fact, we could find another crossover point, $p_{\times_2}^{\prime}$ scaling with $L^{-\eta}$ which the mean total length behaves like $\mathcal{F}[x]\sim x^{\beta}$ for $x\ll 1$, and is a $const$ value for $x\gg 1$. Fig.~\ref{fig::fig3}(b) shows the scaling behaviors for different lattice size of $L$. As it shown, we have a good data collapse with $\eta=0.15\pm 0.02$, this clearly shows that the argument of $p^{\prime}L^{\eta}$ in crossover point should be independent of lattice size, so the crossing probability $p_{\times_2}^{\prime}$ scales like $L^{-\eta}$ with system size. The overlap of the different curves confirms that the fractal dimension of the LERW$_p$ curves above $p_c$ is $\frac{5}{4}$. 
Three different regimes, as shown in Fig. \ref{fig::fig3}, are clearly identified; for $p^{\prime}<p_{\times_1}^{\prime}$ the mean total length scales as $S\sim L^{d_f}$, for $p_{\times_1}^{\prime}<p^{\prime}<p_{\times_2}^{\prime}$, $S$ has a power law behavior as $p^{\prime\beta}$, and finally for $p_{\times_2}^{\prime}<p^{\prime}$, it scales with Euclidean exponent, i.e. $L^{1.25}$. Therefore, the following relation can be obtained, 
\begin{equation}\label{eq::scaling.relation} 
\theta-\eta=\beta^{-1}\left(\frac{5}{4}-d_f\right), 
\end{equation} 
which is in good agreement with our obtained numerical values for the exponents. 
\begin{figure}[b] 
\includegraphics[scale=0.42]{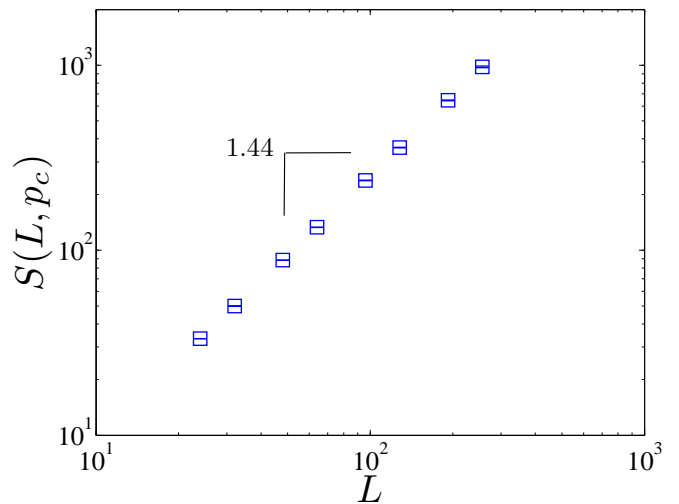} 
\caption{\label{fractal_3D} (Color online) Mean total length of 3D LERW on critical percolation as a function of lattice size $L$. 
Simulations were performed on simple-cubic lattice of size $ L$ ranging from 24 to 256. The number of samples generated for each lattice size ranges from $10^5$ for the smallest system sizes down to about $2\times10^3$ for the largest system sizes. Error bars are smaller than the symbol size.} 
\end{figure}

\emph{Three-dimensional LERW$_p$}.~Our unexpected results of planar LERW$_p$ motivate us to investigate it on the higher dimensions i.e. 3D. The scaling exponents of LERW on Euclidean lattice in three dimensions are not rigorously known. However, the fractal dimension of this model has been reported based on Monte Carlo simulations to be $\approx1.624$ \cite{LERW_3D,Dhar,LERW_Peter}. Since a single exponent would not be enough to describe the scaling behavior of mean total length, $S$, as discussed in\cite{LERW_Peter,*Agrawal01,*Guttmann90}, we cannot study $S(L)/L^{d_f}$ like Fig.\ref{fig::fig2}. In fact, due to finite size effects, corrections to scaling is needed to estimate $d_f$ with more precise results. We restrict our attention to only the problem of LERW on critical percolation cluster. We compute the mean total length $S$ for different lattice size $L$, Fig.\ref{fractal_3D} shows dependence of the $S(L,p_c)$ on lattice size $L$. The fractal dimension of this model is $d_f =1.44 \pm 0.03$. Interestingly, this value is similar to the fractal dimensions of the optimal paths in strongly disordered media in 3D \cite{Cieplak94}.

\emph{Summary and Discussion}.~In this paper, we have investigated regarding the geometrical behavior of LERW$_p$ on percolation cluster, with occupation probability above and equal to the critical value, $p\geq p_c$. Our results show that the scaling behavior of planar LERW$_p$ for $p>p_c$ is the same as the LERW on Euclidean lattices, which has been rigorously proven recently \cite{Yadin11}. However, the LERW on critical percolation clusters scales with an anomalous exponent, our results reveal that the fractal dimension of this model is $d_{f} \approx 1.22$ in 2D. Interestingly, this value is statistically identical to a family of curves appearing in different contexts such as, e.g., polymers in strongly disordered media \cite{Porto97}, watershed of random landscapes\cite{Fehr09,Fehr11,Daryaei12}, bridge percolation \cite{Schrenk12}, minimum spanning tree (MST) \cite{MST}, and optimal path cracks \cite{Andrade09}. 
Also our attempts to understand LERW$_p$ in three dimensions are focused on the critical percolation cluster; the fractal dimension of this model on 3D percolation clusters is $\approx 1.44$. These exponents in both two and three dimensions are similar to the fractal dimensions of the optimal paths in strongly disordered media \cite{Cieplak94}. This fact clearly indicates that optimal paths in strongly disordered media is related to the well-known LERW on critical percolation cluster. Since LERW is equivalent to UST and $q$-state Potts model (in the limit of $q\to 0$), it is interesting to study of these models in a diluted lattice generating by a sequence of random deletions. Because of the negative specific heat exponent of the pure system, Harris criterion \cite{Harris74} claims the universality of this model should remain unchanged. However, as it was reported for spanning trees on critical percolation cluster \cite{Sweeney13}, it is in a different universality class from UST model. Although here we restrict our attention to two and three dimensions, the new findings for spanning trees on critical percolation cluster \cite{Sweeney13} indicate that the upper critical dimension of this model is likely to be the same as percolation model. The LERW on percolation cluster is related to more general concept of random walks on fractal landscape, as another example, it was shown recently that the fractal dimension of LERW on Sierpinski gasket is $\approx1.194$ \cite{Hattori12,Shinoda13}. Finally, our results demonstrate that this model can be classified into two distinct universality classes: LERW on {\it Euclidean} and {\it fractal} geometries.

Near the percolation threshold, $p_c$, there is a crossover regime, shown in Fig.\ref{fig::fig2}, between these two universality classes. To achieve a better understanding of this regime, we have considered the mean total length of LERW$_p$ as a homogeneous function on the lattice size and occupation probability. 
Our findings for crossover regime, shown in Fig.\ref{fig::fig3}, clearly demonstrates that for a fixed lattice size, $L$, three distinct scaling regimes have to be distinguished: (a) a fractal regime for $p^{\prime}<L^{-\theta}$; (b) a Euclidean geometry regime for $p^{\prime}>L^{-\eta}$; (c) a transition regime from the fractal to the Euclidean behaviors for $L^{-\theta}<p^{\prime}<L^{-\eta}$. These regimes are separated by two crossovers. Finally there is a scaling relation between the corresponding crossover exponents and the fractal dimension of LERW. In general, the existence of three different scaling behaviors in a system often leads to three distinct regimes, as it has been reported in both related \cite{Moreira12,Andrade09} and unrelated systems (for example see \cite{Maggs89,*Vacher90,*Konschelle07,*Krusin92,Tokatly04}). Although there are some similar reports in which the crossover behavior has been investigated by just one crossover exponent \cite{Schrenk12,Schrenk12_2,Schrenk13,Nuno11}, the existence of three regimes clearly can be observed in related figures. Moreover, the proposed scaling {\it ansatz} for the systems is not valid in all crossover region.

The connection between LERW$_p$ and other important model such OP and watershed model allows one to look at such random path with a new eye and to build bridges between connectivity in disordered media and other research areas in mathematics, percolation, and quantum field theory. This work opens up several challenges. Besides the need for more precise numerical simulations in higher dimensions to study of the fractal properties and the crossover exponents, it would be interesting to formulate a field theory scheme in fractal landscape. Since continuum limit of the planar LERW on the Euclidean lattice can be descried with Schramm-Loewner evolution \cite{Schramm00}, another interesting possibility is to find a conformal field theory for LERW on critical percolation clusters. 

\begin{acknowledgments} 
We are grateful to N. A. M. Ara\'ujo, H. J. Herrmann, S.Moghimi-Araghi, A. Saberi and M. A. Rajabpour for helpful comments on the manuscript. 
\end{acknowledgments} 

\bibliography{LERW_p} 
\end{document}